\documentclass[sigconf, screen, natbib=false]{acmart}

\usepackage[utf8]{inputenc}
\usepackage[T1]{fontenc}

\usepackage{graphicx}
\usepackage{subcaption}
\usepackage[export]{adjustbox}
\usepackage{textcomp}
\usepackage[hang,flushmargin]{footmisc}
\usepackage{url}
\usepackage{import}
\usepackage{color}

\usepackage{soul}
\usepackage{pifont}
\usepackage{enumitem}
\usepackage{multirow}
\usepackage{blindtext}
\usepackage{dirtytalk} % \say command for quotations
\usepackage{tabularx} % For convenient table columns use column type X
\usepackage{multirow}
\usepackage{listings} % For code snippets
\usepackage{todonotes}
\usepackage{mathtools}
\usepackage{arydshln}
\usepackage{cleveref}
\usepackage{rotating}
\usepackage{balance}
\RequirePackage[
  datamodel=acmdatamodel,
  style=acmnumeric,
  natbib,
  backend=biber,
  mincitenames=1, 
  maxcitenames=1, 
  uniquelist=true,
  giveninits=true
  ]{biblatex}
\bibliography{main}

\newcounter{svcounter}

\graphicspath{ {./Figures/} }
\usepackage[autostyle=true]{csquotes}
\MakeOuterQuote{"}

\lstdefinestyle{python}{ 
    xleftmargin=6.0ex,
    xrightmargin=2.0ex,
    numbers=left,
    frame=single
}
\definecolor{gray10}{gray}{.9}
\definecolor{arsenic}{rgb}{0.23, 0.27, 0.29}
\usepackage{color}

\RequirePackage{fontawesome}

\definecolor{gray50}{gray}{.5}
\definecolor{gray40}{gray}{.6}
\definecolor{gray30}{gray}{.7}
\definecolor{gray20}{gray}{.8}
\definecolor{gray10}{gray}{.9}
\definecolor{gray05}{gray}{.95}

\newlength\Linewidth
\def\findlength{\setlength\Linewidth\linewidth
    \addtolength\Linewidth{-4\fboxrule}
    \addtolength\Linewidth{-3\fboxsep}
}

\newenvironment{rqbox}{\par\begingroup
	\setlength{\fboxsep}{5pt}\findlength
	\setbox0=\vbox\bgroup\noindent
	\hsize=0.95\linewidth
	\begin{minipage}{0.95\linewidth}\normalsize}
	{\end{minipage}\egroup
	\textcolor{gray20}{\fboxsep1.5pt\fbox
		{\fboxsep5pt\colorbox{gray05}{\normalcolor\box0}}}
	\endgroup\par\noindent
	\normalcolor\ignorespacesafterend}

\newcommand{\reviewerOne}[1]{\textcolor{black}{#1}}
\newcommand{\reviewerTwo}[1]{\textcolor{black}{#1}}

\begin{document}

\title{In Search of Metrics to Guide Developer-Based Refactoring Recommendations}
\subtitle{(A Registered Report)}

\author{Mikel Robredo}
   \affiliation{
   \institution{University of Oulu, Finland} 
%   \city{Oulu}
   % \country{Finland}
 }
 \email{mikel.robredomanero@oulu.fi}

\author{Matteo Esposito}
   \affiliation{
   \institution{University of Oulu, Finland} 
%   \city{Oulu}
%   \country{Finland}
 }
 \email{matteo.esposito@oulu.fi}

 \author{Fabio Palomba}
 \affiliation{
   \institution{University of Salerno, Italy} 
 %  \city{Salerno}
%   \country{Italy}
 }
 \email{fpalomba@unisa.it}

 \author{Rafael Pe\~naloza}
 \affiliation{
   \institution{University of Milano-Bicocca, Italy} 
%   \city{Milano}
%   \country{Italy}
 }
 \email{rafael.penaloza@unimib.it}

 \author{Valentina Lenarduzzi}
 \affiliation{
   \institution{University of Oulu, Finland} 
%   \city{Oulu}
%   \country{Finland}
 }
 \email{valentina.lenarduzzi@oulu.fi}

\renewcommand{\shortauthors}{Robredo, et al.}

\begin{abstract}
\textbf{Context}. Source code refactoring is a well-established approach to improving source code quality without compromising its external behavior.
\textbf{Motivation}. 
The literature described the benefits of refactoring, yet its application in practice is threatened by the high cost of time, resource allocation, and effort required to perform it continuously. Providing refactoring recommendations closer to what developers perceive as relevant may support the broader application of refactoring in practice and drive prioritization efforts.
\textbf{Aim}. \reviewerTwo{In this paper, we aim to foster the design of a developer-based refactoring recommender}, proposing an empirical study into the metrics that study the developer's willingness to apply refactoring operations. We build upon previous work describing the developer's motivations for refactoring and investigate how product and process metrics may grasp those motivations.
%Our novel approach promises to deliver the first developer-based refactoring recommendations to effectively target the correct amount of refactoring effort toward the right target.
\textbf{Expected Results}. 
We will quantify the value of product and process metrics in grasping developers' motivations to perform refactoring, thus providing a catalog of metrics for developer-based refactoring recommenders to use.
%Our study will identify metrics to guide developer-based refactoring recommendations. We will provide the first evaluation of LLM, i.e., GPT3.5 and GPT4, in extracting refactoring motivation from commit messages against the motivation ground truth from developers. We will perform open coding to asses metrics that allow the creation of developer-based refactoring recommendations. 
\end{abstract}

\keywords{Empirical Software Engineering, Refactoring, Software Quality}

\maketitle

\section{Introduction}
\label{sec:Intro}
Maintaining code bases in the era of computing pervasiveness in daily tasks, from smart appliances to autonomous vehicles, is getting increasingly challenging and daunting \cite{esposito2023uncovering}. With years of development and increasing reduction of time-to-releases, developers can make poor design choices to meet deadlines, leading to complex and hard-to-maintain software, thus increasing subsequent maintenance effort and costs for software firms \cite{lenarduzzi2021systematic}.
Code refactoring is one of the most known techniques to mitigate software complexity\cite{10.1145/2393596.2393655}. Refactoring aims to introduce code structure changes without altering external behavior, from code optimization to architecture and design patterns \cite{peruma2022refactor}. 

Therefore, developers restructure the code to improve its design, readability, and maintainability to ensure the long-term sustainability of software systems \cite{10.1145/2393596.2393655, LACERDA2020110610,6975643}. 

Thus, developers perceive it as a significant time and resource allocation cost due to its complexity and limitations \cite{10.1145/2393596.2393655,avgeriou2020overview}. Therefore, with limited time-to-release windows, developers often limit resources spent on refactoring, hindering the quality and maintenance of the code \cite{avgeriou2020overview}. 
Nonetheless, reducing software complexity with minor code refactoring can do little to no improvement, leading to decreased code quality. In contrast, bulk and more comprehensive refactoring has a more profound influence on code quality \cite{LACERDA2020110610,6975643}.

Choosing the correct refactoring strategy can be challenging, leading developers to seek assistance. Thus, developers need recommendations to guide their refactoring effort \cite{peruma2022refactor,6975643}. Therefore, our work aims to grasp developers' motivation for refactoring. More specifically, our study identifies metrics to guide developer-based refactoring recommendations. 

No previous study investigated metrics to guide refactoring by exploiting Large Language Models (LLMs) to analyze developers' commit messages and extract the code refactoring motivations. Our novel approach, which departs from conventional techniques, promises to deliver the first developer-based refactoring recommendations to effectively target the correct amount of refactoring effort toward the right target. Our approach benefits practitioners, translating into a cost reduction, prioritizing refactoring efforts, and researchers paving the way for future works in developer-based software maintenance. 

%Our contributions will be:
%\begin {itemize}   
%    \item We will provide the first evaluation of  LLM, i.e., GPT3.5, GPT4, Claude, or Mistral 7B, for instance, in extracting refactoring motivation from commit messages against the motivation ground truth from developers.
%    \item We will asses metrics allowing the creation of developer-based refactoring recommendations.
%\end{itemize}

Finally, our work is based on \citet{silva2016we} and extends \citet{10.1145/3408302}. More specifically, as compared to the latter, the following notable differences can be highlighted for what concerns the study design and goals:

\begin{itemize}
    \item \textbf{Study design: Commit and Pull Requests (PR) Analysis.} While Pantiuchina et al.~\cite{10.1145/3408302} analyzed PR based on keyword lookup and the RMiner tool to detect refactoring-related commits, we propose a \textbf{novel approach based on LLMs} instead of keyword analysis. Moreover, PRs represent developer interaction but can not represent the full project change history \cite{6976151,10.1145/2597073.2597121}. Moreover, PRs can be rejected; thus, it is also challenging to discern expert contributions among PRs accepted and not \cite{YU2016204}. \reviewerTwo{Therefore our work will focus on the use of all refactoring commits and not only on PRs.}
    \item \textbf{Study design: Quality Metrics vs Product and Process Metrics.} Pantiuchina et al.~\cite{10.1145/3408302} focused on product-quality metrics such as  Object Oriented Programming (OOP) complexity, static analysis warnings, and developer-related metrics. We decided to focus on \textbf{established product and process metrics} as suggested by \citet{kamei2012large,rahm2013} as well as the ones introduced by Pantiuchina et al.~\cite{10.1145/3408302}.
    \item \textbf{Goals:} Pantiuchina et al.~\cite{10.1145/3408302} were interested in assessing the \textbf{correlation} between product quality and developer-related metrics and the occurrence of at least one refactor in a specific PR. However, although PR is a richer container of developer thoughts and reflections \cite{10.1145/3408302}, refactoring does not happen only in PR. Therefore, our study focuses on a broader set of metrics analyzing qualitatively and quantitatively the \textbf{motivation} that push developers to refactor to \textbf{define a catalog} of motivations supported by a large number of metrics regardless to the connection of a specific RP.
\end{itemize}

\textbf{Paper Structure}. In Section~\ref{sec:CS}, we present the study design. Section~\ref{sec:Threat} focuses on threats to the validity of our study. Section~\ref{sec:relworks} discusses related work and in Section~\ref{sec:Conclusions}, we draw the conclusions. 
\section{Study Design}
\label{sec:CS}
\begin{figure}[b]
    \centering
    \includegraphics[width=1\columnwidth]{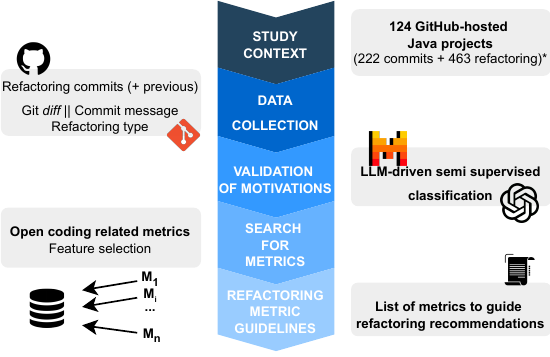}
    \caption{Graphical illustration of the presented study design. \textit{M}: Metric, \textit{i}: Metric type, \textit{n: Total \# of metrics}. (*: Set of analyzed commits providing refactoring motivations in the study from Silva et al.~\cite{silva2016we}).}
    \label{fig:study-design}
\end{figure}

This section outlines the empirical study, including the study goal and research questions, the study context, the data collection methodology, and the data analysis approach.
Our empirical study follows established guidelines defined by Wohlin et al. \cite{Wohlin2000}.  We will publish the raw data in the replication package. Figure~\ref{fig:study-design} provides a graphical description of the study design of the presented plan.

\subsection{Goal and Research Questions}
\label{sec:RQs}
The \emph{goal} of this study is to operationalize the motivations that drive developers to perform refactoring activities, with the \emph{purpose} of identifying product and process metrics that can grasp the motivations pushing developers to perform refactoring. The \emph{perspective} is of researchers and practitioners seeking additional support to guide developers in performing the refactoring. The \textit{context} is open source Java projects. 

% This study aims to identify the motivations that drive developers to perform refactoring activities, those motivations through the use of product and process metrics, validating them with the existing ground truth results~\cite{silva2016we}, and capture the identified motivation through the existing version control metrics in the commit history of projects. 

Therefore, we derived the following research questions (RQs):

\begin{center}
\begin{rqbox}
\textbf{RQ$_1$.} Do motivations for refactoring in past studies align with those found in software project change histories?
%To what extent are the motivations for refactoring identified in previous literature found in refactoring the change history of software projects?
\end{rqbox}
\end{center}

% \begin{center}
% \begin{rqbox}
% \textbf{RQ$_1$.} Are the ground truth questions translated or identified in the refactoring activity?
% \end{rqbox}
% \end{center}
Silva et al.~\cite{silva2016we} proposed a list of motivations for refactoring efforts provided by professional developers.  Pantiuchina et al. \cite{10.1145/3408302} build upon Silva et al.~\cite{silva2016we} motivations focusing on quality metrics and static analysis warnings extracting information from PRs. 
Nonetheless, PRs represent developer interaction but can not represent the full project change history \cite{6976151,10.1145/2597073.2597121}. Moreover, PRs can be rejected; thus, it is also challenging to discern expert contributions among PRs accepted and not \cite{YU2016204}. Hence, we investigate developer motivation, analyzing the single commits with a broader selection of product and process metrics. No prior study considered employing the fully established product and process metrics \cite{kamei2012large,rahm2013} with the refactoring history.

Furthermore, developers' commits, PRs, and code comments may not be able to capture all the motivations pushing developers to refactor the code base; hence, we ask:

%Based on  from the authors survey, 
%We put forward that the refactoring activity in the collected data should demonstrate the validity of the motivations reported by the answers that professional developers provided in the work performed by Silva et al.~\cite{silva2016we}. Therefore, answering this question aims at validating the existing ground truth motivations for driving refactoring activity. 
% We will consider classifying the collected refactoring types based on semi-supervised LLM-driven classification, a process further described in Section~\ref{sec:llm}. 

% \begin{center}
% \begin{rqbox}
% \textbf{RQ$_2$.} Which refactoring types do not match with the stated motivations and what motivation do LLMs relate to them?
% \end{rqbox}
% \end{center}

\begin{center}
\begin{rqbox}
\textbf{RQ$_2$.} Are there additional motivations driving the developers' willingness to perform refactoring?
\end{rqbox}
\end{center}

Refactoring improves source code quality \cite{peruma2022refactor}. Nonetheless, quality improvements in readability, performance, safety, and security, are only some aspects of the bigger picture \cite{7961515}. For instance, in an industrial context, it is essential to assess refactor opportunities in situations with limited resources regarding work allocation and time \cite{10.1007/978-3-319-09156-3_37}. Therefore, we investigate whether external limitations or additional motivation influence refactoring efforts. 

Finally, extending Pantiuchina et al., \cite{10.1145/3408302} metric assessment, we ask:

%In the possible context of identifying further motivations to perform the considered refactoring types and therefore presenting discrepancies with the ground truth, this question aims to add the underlying motivations for refactoring activity and thus augment the knowledge within the ground truth motivations.

% \begin{center}
% \begin{rqbox}
% \textbf{RQ$_3$.} Can we capture the identified motivation in the version control history metrics?
% \end{rqbox}
% \end{center}

\begin{center}
\begin{rqbox}
\textbf{RQ$_3$.} To what extent can product and process metrics capture the motivations driving the developers' willingness to perform refactoring?
\end{rqbox}
\end{center}

% To what extent can product and process metrics capture the motivations driving the developers' willingness to perform refactoring?

Based on  \citet{10.1145/3408302}, a relationship exists between specific quality metrics and refactoring operations. Nonetheless, \citet{10.1145/3408302} metric selection was limited to quality metrics and static analysis warnings. We extended their selection with established product and process metrics \cite{kamei2012large,rahm2013}  thus broadening the candidate for a plausible relationship to refactoring motivations. 
%Matteo: quali sono e perche non le usiamo ma usiamo altre
Therefore, we hypothesize that there must be already existing metrics in the commit history of software projects highly correlated to the ground truth motivations and the hypothetically newly identified ones. The formulated hypotheses are defined as follows: 

\begin{itemize}
    \item \textbf{H}$_0$ \reviewerOne{There is no correlation between the considered metrics and refactoring motivations.}
    \item \textbf{H}$_1$ \reviewerOne{There is no correlation between the considered metrics and refactoring motivations.}
\end{itemize}

\reviewerOne{Consequently, considering \emph{H$_1$} as the accepted hypothesis and therefore the evidence of correlation, we expand our hypotheses as follows:}

\begin{itemize}
    \item \textbf{H}$_{1.1}$ \reviewerOne{The identified correlation shows a negative direction.}
    \item \textbf{H}$_{1.2}$ \reviewerOne{The identified correlation shows a positive direction.}
\end{itemize}

In this sense, this question aims at identifying the metrics that can explain the existence of the stated motivations \reviewerOne{, and further understand their specific implications}. And therefore provide a list of metrics to monitor and guide developers for refactoring recommendations.

\subsection{Study Context}
\label{sec:Context}

Our study context is based on Silva et al.~\cite{silva2016we} dataset. Silva et al.~\cite{silva2016we} initially selected the top 1000 Java repositories in GitHub~\footnote{https://github.com}, ordered in terms of popularity, and further filtered it, removing the lower quartile based on the number of commits in those projects, thus obtaining a list of 748 repositories with consistent maintenance activities.
The dataset considered in Silva et al.~\cite{silva2016we} study was reduced to 471, filtering out those not in active development while writing their paper.
%Furthermore, as they were interested in projects that were active during the study, they reduced the number of repositories to 471, which were the only ones that presented one commit during the study period. 
The final set of considered projects comprised 124 projects, which featured at least one refactoring during the study period and an answer provided by the developers of the specific refactoring activity to the author's questions on the refactoring motivation.
The final set of projects comprised 222 commits, for which the authors provided answers regarding their motivation for performing the refactoring. Within this set of commits, 463 refactorings from 12 refactoring types were detected using RefactoringMiner~\cite{tsantalis2013multidimensional} tool. 
Therefore, to validate the motivations provided by developers, our study considers the final set of 124 GitHub-hosted Java projects described in the dataset~\footnote{http://aserg-ufmg.github.io/why-we-refactor} published by Silva et al.~\cite{silva2016we}.

\subsection{Data Collection}
\label{sec:DataCollection}

In the first step of the data collection, we will gather all the metrics we considered in our study presented in Table~\ref{tab:jit-metrics} from the 124 repositories of the adopted dataset~\cite{silva2016we}. Although the dataset already provides the \textit{True Positive} commits with refactoring, we will run the RefactoringMiner~\cite{tsantalis2013multidimensional} tool to detect further refactoring activity performed until now. We aim to follow the same collection procedure Silva et al.~\cite{silva2016we} performed for consistency in our results. Therefore, we expect to identify commits in which the refactoring activity was performed from the time the previous study was performed as well. Next, we will collect the commit message of those commits in which refactoring activity was detected and further collect their preceding commit. Thus, we aim to capture the change in the code from the commit that preceded the refactoring activity, the refactoring type itself, and the message that details the performed change in the project's version control history.

%in their study only 471 projects resulted in being active at that time, we will initially consider all of them for our study, which can consider refactoring activity that occurred in the past.

%Therefore, we aim at collecting the existing data published in the previously mentioned study, but will also run the RefactoringMiner~\cite{tsantalis2013multidimensional} tool with the remaining inactive projects following the same procedure they did for the sake of consistency in our results. In this way, we expect to identify the commits in which the refactoring activity of the considered refactoring types was performed. 
%Consequently, we will collect the commit message of the commits with detected refactoring activity and further collect their preceding commit. Thus, we aim to capture the change in the code from the commit that preceded the refactoring activity, the refactoring type itself, and the message that informed us about the performed change in the project's version control history.

% \begin{figure}[h!]
%     \centering
%     \includegraphics[width=0.8\columnwidth]{Figures/vertical-refactoring-example.pdf}
%     \caption{Input example of Extract Method refactoring and commit message from project \textit{elastic/elasticsearch}~\cite{githubGitHubElasticelasticsearch}.}
%     \label{fig:refactoring-example}
% \end{figure}

We will consider the motivations reported by Silva et al.~\cite{silva2016we}. The list obtained from the motivations reported by developers consists of 44 motivations related to the 12 types considered in the cited study. Table~\ref{tab:refactorings} shows the list of refactoring types, their description, and the number of motivations identified per refactoring. Moreover, Table~\ref{tab:motivations} presents the \textit{label} or brief denomination of the motivations defined by human developers per each of the selected refactoring types. We are not expanding the description concerning motivations due to space constraints in this registered report.

To answer \textbf{RQ$_3$}, we will mine and open code the variables that better describe the resulting motivations obtained from the previous research questions. We will provide further details of the process in Section~\ref{sec:open-coding}.

\begin{table*}[ht]
\footnotesize
% \resizebox{\textwidth}{!}{
\begin{tabular}{lccp{11cm}} 
 \hline
 \textbf{Refactoring Type} & \textbf{Id} & \textbf{\# Motivations} & \textbf{Definition} \\
 \hline
 EXTRACT METHOD & EM & 11 & Takes a clump of code and turns it into its method.\\ 
 MOVE CLASS & MC & 9 & Move the class to its new folder on the source tree. \\
 MOVE ATTRIBUTE & MA & 2 & All attributes matching the selected attribute name on tags with the selected tag name may be moved inward toward a subtag of a given name. \\
 RENAME PACKAGE & RP & 3 & Renames the name of the selected project package. \\
 MOVE METHOD & MM & 5 & Creates a new method with a similar body in the class it uses most. Either turns the old method into a simple delegation or remove it altogether. \\
 INLINE METHOD & IM & 3 & Puts the method's body into the body of its callers and removes the method. \\
 PULL UP METHOD & PUM & 1 & Moves methods with identical results on subclasses to the superclass. \\
 PULL UP ATTRIBUTE & PUA & 1 & Moves attributes with identical results on subclasses into the superclass. \\
 EXTRACT SUPERCLASS & ES & 3 & Creates a superclass and moves the common features to the superclass. \\
 PUSH DOWN METHOD & PDM & 1 & Given a field only used by some subclasses, it moves the field to those subclasses. \\
 PUSH DOWN ATTRIBUTE & PDA & 2 & Given an attribute only used by some subclasses, it moves the attribute to those subclasses. \\
 EXTRACT INTERFACE & EI & 3 & Given two classes having part of their interfaces in common, it extracts the subset into an interface. \\
 \hline
\end{tabular}
\caption{Definitions of the considered refactoring types~\cite{fowler1999refactoring}}
\label{tab:refactorings}
\end{table*}

\begin{table*}[ht]
\footnotesize
% \resizebox{\textwidth}{!}{
\begin{tabular}{l|p{4.5cm}lp{4cm}} 
 \hline
 \textbf{Refactoring Type} & \textbf{} & \textbf{Motivations} & \textbf{} \\
 \hline
 EXTRACT METHOD & Extract reusable method & Introduce alternative method signature & Decompose method to improve readability\\
  & Facilitate extension & Remove duplication & Replace Method preserving backward compatibility\\
  & Improve testability & Enable overriding & Enable recursion\\
  & Introduce factory method & Introduce async operation & \\
  
 MOVE CLASS & Move class to appropriate container & Introduce sub-package & Convert to top-level container\\
  & Remove inner classes from deprecated container
 & Remove from public API & Convert to inner class\\
  & Eliminate dependencies & Eliminate redundant sub-package & Backward compatibility\\
  
 MOVE ATTRIBUTE & Move attribute to appropriate class & Remove duplication & \\ 
 
RENAME PACKAGE & Improve package name & Enforce naming consistency & Move package to appropriate container\\ 
 
 MOVE METHOD & Move method to appropriate class & Move method to enable reuse & Eliminate dependencies\\ 
  & Remove duplication & Enable overriding & \\
 
 INLINE METHOD & Eliminate unnecessary method & Caller becomes trivial & Improve readability\\ 
 
 PULL UP METHOD & Move up common methods &  & \\ 
 
 PULL UP ATTRIBUTE & Move up common attributes &  & \\ 
 
 EXTRACT SUPERCLASS & Extract common state/behavior & Eliminate dependencies & Decompose class\\ 
 
 PUSH DOWN METHOD & Specialized implementation &  & \\ 
 
 PUSH DOWN ATTRIBUTE & Specialized implementation & Eliminate dependencies & \\ 
 
 EXTRACT INTERFACE & Facilitate extension & Enable dependency injection & Eliminate dependencies\\ 
 \hline
\end{tabular}
\caption{Motivations derived from developers' answers based on their refactoring activity}
% \todo[inline]{Mikel@Fabio: Prendiamo anche delle product metrics da Rahman?}
\label{tab:motivations}
\end{table*}

\subsection{Data Analysis}
\label{sec:DataAnalysis}
% This section describes the data analysis process and the techniques and methods for answering our RQs. 

\subsubsection*{Refactoring motivation identification through refactoring commit activity (RQ1)}
\label{sec:llm}
We are interested in grasping the motivations that led the developers to perform the selected refactoring types and further compare if these motivations are related to the ones resulting from the previous work. 
We plan to perform a \textit{multi-class semi-supervised} classification of the collected refactoring commits based on the identified motivations to achieve this.
The classification process will be rooted in the use of different LLMs; we plan to use the APIs of LLMs such as chatGPT3.5 or chatGPT4\footnote{https://openai.com/product} following the state-of-the-art~\cite{chang2023survey} along with further models such as Mistral 7B\footnote{https://www.lemonfox.ai/mistral-api} for instance, which has shown promising performance as well~\cite{jiang2023mistral}. Since we already have supervised answers for a set of 222 commits with their respective 463 refactorings from the study context, we will set our LLMs to learn with this supervised training data, following the learning paradigm called \textit{Few-Shot Learning} (FSL)~\cite{wang2020generalizing, fei2006one, fink2004object}. This initial step will provide the commit change between the commit in which the refactoring has been implemented and its preceding commit, the commit message, and the refactoring type and motivation involved in each case from the supervised data. Thus, the LLMs will be initially provided with the motivations and refactoring types detailed in Tables~\ref{tab:refactorings} and~\ref{tab:motivations} as the baseline framework for the classification approach. Then, to analyze the motivation behind each of the collected refactoring commits, we will fit the LLMs with the commit change between the refactoring commit and the preceding commit. Additionally, we will provide the message provided in both commits. We will consider fitting the models only with the described content and a defined explanatory question. 
%Figure~\ref{fig:refactoring-example} provides an input example of refactoring activity for the type Extract Method and the message provided by the committer.
We will consider further motivations LLMs suggest when discrepancies are found between the provided refactoring cases and the existing refactoring motivations since we are also interested in discovering potential motivations behind the refactoring activity that may not have been addressed in previous research. Thus, we aim to analyze possible deviations from the developers' provided motivations in the real code.

\subsubsection*{Comparison of the resulting motivations (RQ2)}
\label{sec:comparison}

We are interested in analyzing developers' accuracy when providing their main motivations to perform a refactoring activity, considering the motivations that the LLMs have extracted from the provided commit changes and messages. 
Therefore, in this analysis stage, we will compare the frequency difference between the reported AI-based motivations and those obtained in~\cite{silva2016we} for each of the considered refactoring types.
Similarly, since we have allowed the LLMs to provide new motivations when finding discrepancies between the existing motivations and the refactoring case, we want to compare the difference between the initially considered motivations and those that LLMs relate to for the same refactoring type.
% \todo[inline]{Mikel: Feedback for RQ1 analysis and RQ2 analysis}

\subsubsection*{Capturing motivation metrics in version control history (RQ3)}
\label{sec:open-coding}

To answer our third research question, the current section of the data analysis aims to capture the software metrics existing in the considered projects that better describe the scenario that motivates the developer to perform the refactoring activity and further test their relationship through the defined hypotheses in Section~\ref{sec:RQs}.

To accomplish that, we will initially open code the selected software projects and collect software metrics when each refactoring was performed at the project stage. We will consider the software metrics that are, by definition, related to the motivations described both by developers~\cite{silva2016we} and the ones reported by the LLMs. 
\begin{table*}[h]
\centering
\scriptsize
\begin{tabular}{l|l|p{15.7cm}}
\hline
& \textbf{Metric} & \textbf{Description} \\ 
\hline 
\multirow{15}{*}{\rotatebox[origin=c]{90}{Rahman~\cite{rahm2013}}} 
& COMM     & The cumulative number of changes in a given file up to the considered commit.                                                  \\
& ADEV     & The cumulative number of active developers who modified a given file up to the considered commit.                              \\
& DDEV     & The cumulative number of distinct developers contributed to a given file up to the considered commit.                          \\
& ADD      & The normalized number of lines added to a given file in the considered commit.                                                 \\
& DEL      & The normalized number of lines removed from a given file in the considered commit.                                             \\
& OWN      & The value indicates whether the file owner does the commit.                                                            \\
& MINOR    & The number of contributors who contributed less than 5\% of a given file up to the considered commit.                          \\
& SCTR     & The number of packages modified by the committer in the considered commit.                                                     \\
& NADEV    & The number of active developers who changed any of the files involved in the commits where the given file has been modified.   \\
& NDDEV    & The number of distinct developers who changed any of the files involved in the commits where the given file has been modified. \\
& NCOMM    & The number of commits where the given has been involved.                                                                       \\
& NSCTR    & The number of different packages touched by the developer in commits where the file has been modified.                         \\
& OEXP     & The percentage of code lines authored by a given developer in the project.                                               \\
& EXP      & The mean of the experience of all developers across the project.                                                         \\
\hline 
\multirow{14}{*}{\rotatebox[origin=c]{90}{Kamei~\cite{kamei2012large}}} 
& ND       & The number of directories involved in a commit.                                                                                \\                                       
& NS       & Number of modified subsystems.                      \\
& NF       & Number of modified files. \\
& ENTROPY  & The distribution of the modified code across each given file in the considered commit.                                         \\
& LA       & Ten lines added to the given file in the considered commit (absolute number of the ADD metric).                      \\
& LD       & The number of lines removed from the given file in the considered commit (absolute number of the DEL metric).                  \\
& LT       & The number of lines of code in the given file in the considered commit before the change.                                      \\
& FIX      & Whether or not the change is a defect fix. \\
& NDEV     & The number of developers that changed the modified files. \\
& AGE      & The average period between the last and the current change.                                                                 \\
& NUC      & The number of times the file has been modified up to considered commit.                                                  \\
& CEXP     & The number of commits performed on the given file by the committer up to the considered commit.                                \\
& REXP     & The number of commits performed on the given file by the committer in the last month.                                          \\
& SEXP     & The number of commits a given developer performs in the considered package containing the given file.                   \\
\hline 
\multirow{19}{*}{\rotatebox[origin=c]{90}{Pantiuchina~\cite{10.1145/3408302}}} 
 & CBO & Coupling Between Object classes: measures the dependencies a class has. \\
 &         WMC & Weighted Methods per Class: sums the cyclomatic complexity of the methods in a class. \\
  &        RFC & Response For a Class: the number of methods in a class plus the number of remote methods that are called recursively [citeme]. \\
 &         ELOC & Effective Lines Of Code: the lines of code excluding blank lines and comments. \\
 &         NOM & Number Of Methods in a class. \\
 &         NOPM & Number Of Public Methods in a class. \\
 &         DIT & Depth of Inheritance Tree: the length of the path from a class to its farthest ancestor.  \\
 &         NOC  & Number Of Children (direct subclasses) of a   class. \\
 &         NOF & Number Of Fields declared in a class. \\
 &         NOSF & Number Of Static Fields declared in a class. \\
 &         NOPF & Number Of Public Fields declared in a class. \\
 &         NOSM & Number Of Static Methods in a class. \\
 &         NOSI & Number Of Static Invocations of a class. \\
 &         HsLCOM & Henderson-Sellers revised Lack of Cohesion   Of Methods (LCOM): a class cohesion metric based on sharing local instance variables by the class methods.\\ %HsLCOM  addresses limitations of the original LCOM. 
 &     C3    & Conceptual Cohesion of Classes: avg. textual similarity between all pairs of methods in a class.  \\
 & StrRead & Structural readability: uses structural aspects (e.g., line length) to model code  readability.  \\
 & ComRead & Comprehensive readability model: combines structural, visual (e.g., alignment),   and textual features (e.g., comments readability).  \\
        
\hline 
\end{tabular}
\caption{Product and process metrics adopted in this study.}
\label{tab:jit-metrics}
\end{table*}

As initial potential metrics, we will consider diverse software metrics such as process metrics defined by Rahman \& Devanbu~\cite{rahm2013} as well as product metrics defined by Kamei et al.~\cite{kamei2012large}. Similarly, we will consider additional metrics used in previous work by Pantiuchina et al. ~\cite{10.1145/3408302}. 
%As initial potential metrics, we will consider software metrics already adopted in fault prediction-based studies~\cite{lomio2021fault} as well as change metrics for quality assurance prediction, where these metrics were used as prediction features~\cite{kamei2012large}. The considered metrics and their concise description can be found in Table~\ref{tab:jit-metrics}.
Then, we will rank the selected metrics by how related or informative they are toward the identified motivations. For that, we will consider two main different approaches. The first approach will consist of quantifying the \textit{feature importance}~\cite{sammut2011encyclopedia} provided by each of the considered metrics in explaining each refactoring motivation accordingly. Feature importance measurement stands as a feature selection technique commonly used in previous research, as demonstrated by Rajbahadur et al.~\cite{9347823}, and can be calculated in different ways depending on the model chosen, e.g., Random Forest or Extreme Gradient Boost, for instance. To validate our results, we will consider a minimum of two models to calculate the importance level.
Similarly, to statistically validate the obtained results from the approach below, we will statistically test the correlation level of the software metrics toward each of the refactoring motivations and thus test the hypotheses defined previously in Section~\ref{sec:RQs}. Since we do not know \textit{a priori} the exact distribution of the considered metrics, we will use the parametric test \textit{Shapiro-Wilk}~\cite{gentle2009computational} and the non-parametric \textit{Kolmorov-Smirnov}~\cite{DeGroot2019} test to test the normality distribution assumption. Based on the obtained results we will consider using parametric tests such as \textit{Pearson's r} correlation coefficient~\cite{cohen2009pearson} or non-parametric tests such as \textit{Spearman's $\rho$} or \textit{Kendall's $\tau$}~\cite{Wohlin2000} coefficient for instance. The proposed tests provide diverse ranked \textit{test statistic} values to measure the extent of correlation of the tested metrics and the statistical significance of the obtained statistic in terms of the well-known \textit{p-value}. Since we are implementing multiple tests in multiple variables simultaneously, which can increase the likelihood of obtaining false statistically significant outcomes, we will also perform the \textit{Bonferroni post-hoc} test~\cite{DeGroot2019} to adjust the obtained \textit{p-values} for the test statistics to the multiple comparison scenario, and thus confirm our formulated hypotheses.
% \todo[inline]{Mikel@Rafael: Feedback per favore}

% \subsection{Empirical Validation}
% \label{sec:Validation}
% \todo[inline]{@vale}

% \subsection{Replicability}
% \label{sec:Replicability}
% To allow the replication of our study, we will publish the raw data in the replication package. 

% \input{Section/Results.tex}
% \input{Section/Discussion.tex}
\section{Threats to Validity}
\label{sec:Threat}
% In this Section, we discuss the threats to validity, including internal, external, construct validity and reliability. 
% We also explain the different adopted tactics~\cite{YinCaseStudies2009}. 

\textbf{Construct Validity}. We acknowledge that using LLMs for \emph{semi supervised multi-class} classification may stand as a threat to the results of this study as it assumes to be correct the artificial intelligence-based feedback. However, to minimize this threat we consider using state-of-the-art as well as well-valued LLMs such as chatGPT and Mistral 7b for instance. Similarly, Silva et al~\cite{silva2016we} already considered the threat of using RefactoringMiner for mining the commits in which the refactorings were performed as some classifications may be missed. However, 
we still consider using RefactoringMiner to provide consistency to our results as one of our goals is to validate the ground truth motivations provided by developers.

\textbf{Internal Validity.} One of the main outcomes of this study will be the publication of a set of metrics to guide developers when performing refactoring activity. These metrics will be based on the human-based motivations as well as AI-based motivations obtained from the performed classification, hence we understand as a threat the fact that the defined refactoring guideline metrics may not provide an accurate instruction to commit refactoring. Missing motivations from the developers as well as potential motivations yet to be discovered by analyzing further projects may contribute to the possible lack of accuracy. However, we understand that the motivations provided by human developers~\cite{silva2016we} provide the closest feedback to the real motivations for committing refactoring, and therefore reduce this type of threat.

\textbf{External Validity.} This study only considers open source projects, based on Java programming languages and hosted by GitHub. Therefore, the results of this study, even though they aim to cover a wide range of projects, cannot claim to apply to systems outside the open-source community or projects developed in a different programming language. However, the presented study plan aims to analyze all the existing refactoring activity of 124 projects for 12 refactoring types which, if the results provide a clear guideline of metrics, would provide one of the most consistent recommendation guidelines for refactoring activity in the field.  

\textbf{Reliability.} The planned data analysis is presented in a format that aims at providing answers to both the refactoring cases in which the motivation is identifiable as well as when discrepancies may exist and therefore already considered motivations may not fit for the refactoring case. Similarly, the choice of statistical tests detailed in this article aims at covering the possible statistical assumptions that data differently distributed may require for performing statistical testing. The source data provides a considerable set of GitHub-hosted projects, yet a small portion of samples given the existing dimension of projects published in the mentioned platform, therefore we assume as a threat the hypothetical differences in results if the same analysis were implemented in a different sample of projects. However, the validation of the developer-based motivations through the results of this study would minimize the presented threat as we understand that the motivations presented by developers are not the result of their performance in the analyzed projects but the result of their entire development career and hence the product of a larger number of projects.
\section{Related Work}
\label{sec:relworks}
Software refactoring can improve different aspects of software quality, such as readability and maintainability \cite{8456513}. Anyhow,  refactoring is also a time-consuming task and, as such, should be correctly prioritized \cite{10.1145/3387940.3392191}.  Therefore, to match practitioners' needs, we must narrow the refactoring scope via recommendations \cite{8477161}.
\citet{8456513} investigated software refactoring, introducing a novel approach that evaluates refactoring solutions based on traceability and source code smell using entropy-based and traditional metrics, respectively. The authors found that their approach outperformed traditional metric-based recommendations in 71\%  of the examined cases.
\citet{kaur2021brief} introduced a multi-objective optimization technique to generate refactoring solutions that maximize the software quality, use of smell severity, and consistency with class importance. They addressed the limitations of existing approaches, which focus solely on software enhancing traditional software quality. Using a multi-objective spotted hyena optimizer (MOSHO), they evaluated refactoring solutions on five open-source datasets. The authors showed MOSHO's effectiveness in leveraging class importance and smell severity scores.

\citet{10.1145/3387940.3392191} proposed a novel approach for recommending Move Method refactoring, mainly useful when methods rely heavily on external class members. Existing approaches, often heuristic-based, suffer from limitations such as metric dependency and manual threshold setting. On the other hand, the proposed approach leverages code2vec, a path-based code representation capturing syntactic and semantic information, to train a machine learning classifier. The empirical study on open-source projects and a synthetic dataset showed that the proposed approach outperformed existing tools like JDeodorant and JMove.

Similarly, \citet{8477161} proposed a dynamic and interactive refactoring recommendation approach to address the challenges of managing software complexity and facilitating software evolution. The proposed approach leverages NSGA-II \cite{DEBUCK2019613} to suggest refactorings while considering developer feedback, aiming to improve software quality while minimizing deviation from the initial design. They evaluated the approach on open-source and industrial projects, demonstrating its superiority over existing search-based techniques and fully automated tools.
Our work does not focus only on the mere refactoring effort but also on the developer's perspective of refactoring; more specifically, we investigate the motivation that leads developers to refactor specific portions of the code base.

Similarly, \citet{10.1145/3408302} performed a large-scale study to understand why developers refactor code in open-source projects, complementing previous findings from developer surveys. They analyzed 287,813 refactoring operations across 150 systems, examining the relationship between refactoring operations and process/product metrics.  The authors highlighted that there exists a relationship between metrics and refactoring operations, along with a detailed taxonomy of refactoring motivations. 
Our study differs from Pantiuchina et al.~\cite{10.1145/3408302} approach in study design and goals. While ~\cite{10.1145/3408302} relied on keyword lookup and the RMiner tool for PR analysis, we propose a novel methodology employing Language Model analysis. Moreover, while ~\cite{10.1145/3408302} focused on product quality and developer-related metrics, our study prioritizes established product and process metrics. Lastly, while the goal was to correlate product quality and developer metrics with refactoring in PRs, we aim to comprehensively analyze motivations for refactoring across various contexts, defining a catalog of motivations supported by a broader selection of metrics from established guidelines.
\section{Conclusion}
\label{sec:Conclusions}
We aim to search for existing metrics to guide practitioners with developer-based refactoring recommendations. In this sense, this study will first validate ground truth motivations driving refactorings provided by developers and similarly will identify further potential motivations through the use of LLMs by analyzing multiple refactoring commits and their messages. 
The second but not least important outcome of this study will be the identification of the most informative metrics for the concerning motivations. This study will validate the list of motivations that drive developers to perform the considered refactoring types and further will add a valuable contribution to the existing body of knowledge in refactoring research, as it will provide a set of metrics directed to practitioners to guide them when performing refactoring in their projects. Moreover, with the obtained metrics we aim to study time-dependent estimation of refactoring proneness, thus providing a concept-temporal refactoring recommendation to practitioners.

%\bibliographystyle{ACM-Reference-Format}
%\bibliography{sample-bibliography}
 \balance
\printbibliography
\end{document}